\documentclass[twocolumn,showpacs,preprintnumbers,amsmath,amssymb]{revtex4}

\usepackage{graphicx}
\usepackage{dcolumn}
\usepackage{bm}

\begin{document}

\title{Tensor network method for solving the Ising model with a magnetic field}
\author{Myung-Hoon Chung\footnote{$E$-$mail$ $address$: mhchung@hongik.ac.kr}} 
\affiliation{College of Science and Technology, Hongik University,
Sejong 339-701, Korea}

\date{\today}

\begin{abstract}

We study the two-dimensional square lattice Ising ferromagnet and antiferromagnet
with a magnetic field by using tensor network method.
Focusing on the role of guage fixing, we present the partition function in terms of a tensor
network. The tensor has a different symmetry property for ferromagnets and antiferromagnets. 
The tensor network of the partition function is interpreted as a multiple product
of the one-dimensional quantum Hamiltonian.
We perform infinite density matrix renormalization group
to contract the two-dimensional tensor network.
We present the numerical result of magnetization and entanglement entropy 
for the Ising ferromagnet
and antiferromagnet side by side.
In order to determine the critical line in the parameter space of temperature and magnetic field,
we use the half-chain entanglement entropy of the one-dimensional quantum state.
The entanglement entropy precisely indicates the critical line forming the parabolic shape
for the antiferromagnetic case, but shows the critical point for the ferromagnetic case.

\end{abstract}

\pacs{71.27.+a, 02.70.-c, 03.67.-a}


\maketitle

\section{INTRODUCTION}

Numerical methods using computers are essential tools for dealing with quantum many-body systems. Since the computational complexity increases exponentially with the size and dimension of the many-body system, the development of numerical methods capable of effectively addressing the problem is essential. In the 1990s, exact diagonalization and quantum Monte Carlo methods were used and since then, tensor network technology \cite{Orus1} has been studied as a basic tool for numerical research, with the aim of efficiently dealing with many-body systems.

Tensor networks \cite{Orus2} have been independently discovered in various fields. The exactly solvable Baxter models of statistical physics \cite{Baxter}, the first exact short-range spin-chain ground state proposed by Affleck, Kennedy, Lieb, and Tasaki \cite{Affleck} in the field of quantum many-body systems, include the concept of tensor networks. On the other hand, White \cite{White} proposed a new algorithm, the density matrix renormalization group (DMRG), and achieved remarkable results in accurately capturing the ground state of large quantum systems. Shortly thereafter, \"{O}stlund and Rommer \cite{Ostlund} showed that the ground state obtained by the DMRG algorithm is exactly the matrix product state (MPS). Nishino and Okunishi \cite{Nishino} also integrated the DMRG into Baxter physics. When the DMRG began to be understood from the perspective of entanglement, a concept of quantum information, the DMRG was completely redeveloped as the matrix product operator (MPO) and MPS \cite{Schollwoeck}. The infinite density matrix renormalization group (iDMRG) algorithm  \cite{McCulloch} is a highly successful numerical algorithm for the study of low-dimensional quantum systems.

Entanglement \cite{Eisert} can be used to describe the interaction relationship between particles in multiple systems, and in fact, tensor networks can be used to represent entangled states concisely and efficiently. The degree of entanglement of the quantum state is often expressed by a single number of entanglement entropy. It is shown that entanglement entropy acts as a marker to detect quantum phase transitions \cite{Cha}.

The one-dimensional Hubbard model \cite{Chung21} and the two-dimensional Ising model \cite{Chung23} have become a prototypical playground for theoretical studies of quantum correlations and classical statistical mechanics.
For the Hubbard model, the so-called doubling the MPO \cite{Chung} was introduced in a similar fashion of fermionic matrix product states \cite{Bultinck}. The partition function for the Ising model is expressed by a two-dimensional tensor network, which was contracted by the iDMRG \cite{Chung24}.

There are many well-known methods to contract two-dimensional tensor networks in two different ways: coarse-graining and boundary handling \cite{Fishman}. For the coarse-graining methods, we list the names:
tensor renormalization group \cite{Levin}, higher-order tensor renormalization group \cite{Xie}, and tensor network renormalization \cite{Evenbly}. In addition, an MPS is used as the ansatz for the boundary, and this MPS is optimized in various ways. The boundary methods include
variational uniform matrix product states \cite{Zauner}, corner transfer matrix renormalization group \cite{Corboz}, and iDMRG.

In this paper, we consider simultaneously the ferromagnetic \cite{Onsager} 
and antiferromagnetic \cite{Muller} Ising models
in the tensor network formalism. The four-legged tensor corresponding to the partition function
has complete symmetry for the ferromagnetic case, but less symmetry for the
antiferromagnetic case. By the iDMRG method, we evaluate the magnetization given by
a defect-containing tensor network,
and the entanglement entropy of the one-dimensional quantum state
related to the one-dimensional transfer operator defined by the partition function \cite{Zhang}.
The entanglement entropy successfully determines the critical line in the parameter space of temperature and external magnetic field. We note that the entanglement entropy 
exhibits a singularity more clearly than the magnetization in the antiferromagnetic Ising model.

\section{Tensor network representation for the partition function}

We examine how a partition function for a statistical model
is represented as a tensor network.
An example of the corresponding prototype is the Ising model.
We emphasize that the tensor network has a different form in the two cases 
of ferromagnetic and antiferromagnetic Ising model.

\begin{figure}
\includegraphics[width= 7.7 cm]{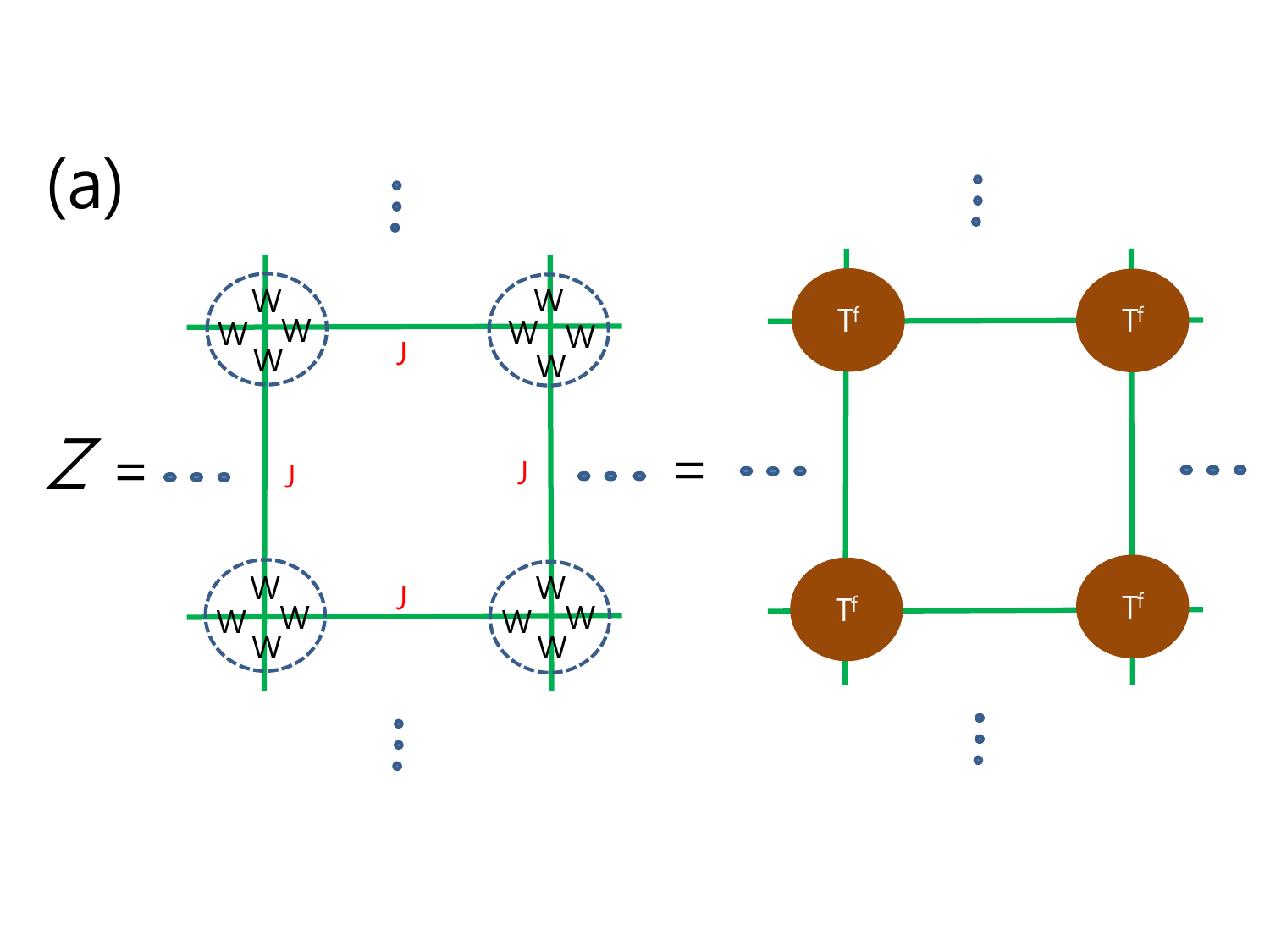}
\includegraphics[width= 7.7 cm]{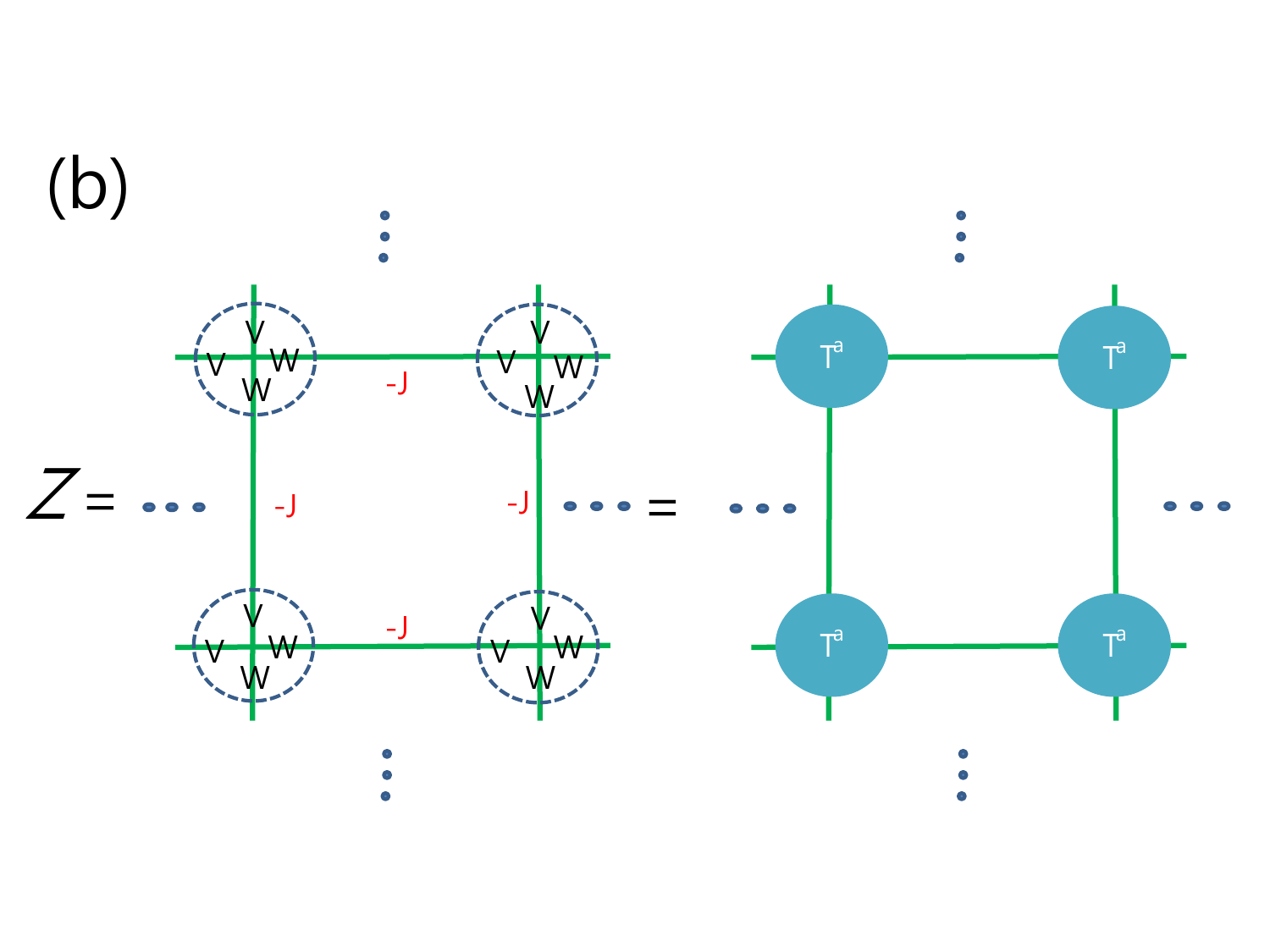}
\includegraphics[width= 7.7 cm]{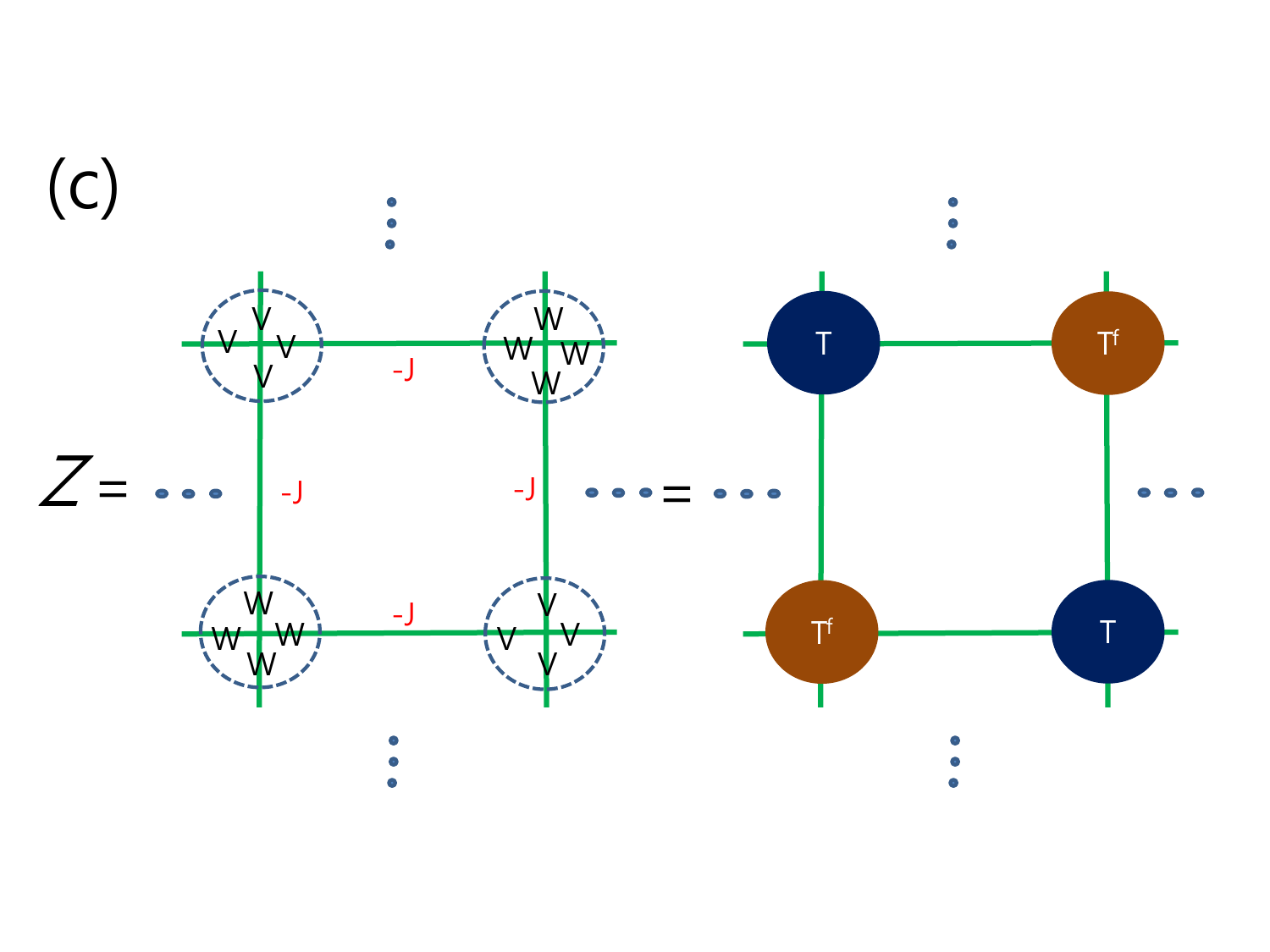}
\caption {The partiton function is written as a tensor network, where the four-legged 
tensor $T_{ruld}^{\mbox{f}}$ and $T_{ruld}^{\mbox{a}}$ with the four indices, right-up-left-down (ruld), is used for the two cases: (a) ferromagnets and (b) antiferromagnets.
By fixing a gauge, (b) the perfect uniform tensor network can be transformed into 
(c) the checkerboard network containing the two constitutive tensors 
realized with 4 $W$ and 4 $V$.
\label{fig:fig1} }
\end{figure}

By definition, the partition function of the Ising model is given by the sum of all
possible spin configurations $\{\sigma\}$:
\begin{equation}
Z = \sum_{\{\sigma\}} e^{-\beta H(\{\sigma \})} ,
\label{eq:Z}
\end{equation}
where $\beta$ is the inversed temperature. For the ferromagnetic (f) and antiferromagnetic (a) cases, the energy $H$ with a magnetic field $B$ and a coupling constant $J>0$ is written as 
\begin{eqnarray}
H^{\mbox{f}} = -J \sum_{\langle i j \rangle}  \sigma_i \sigma_j   - B \sum_{i}  \sigma_i,  \\
H^{\mbox{a}} = J \sum_{\langle i j \rangle}  \sigma_i \sigma_j   - B \sum_{i}  \sigma_i, 
\end{eqnarray}
where $\langle i j \rangle$ represent the nearest-neighbor lattice sites, and
the value of $\sigma$ will be $-1$ or $1$. 
Using the singular value decomposition for the $2 \times 2$ matrix,
we obtain the component tensor $W$ and $V$ for the partition function, such as
\begin{equation}
 e^{\beta J \sigma_i \sigma_j } \rightarrow 
 \left( \begin{array}{cc} 
  e^{\beta J}  &  e^{-\beta J}  \\
 e^{-\beta J}  &  e^{\beta J} 
\end{array} \right)
= \sum_{m=1,2}W_{\sigma_i m}W_{\sigma_j m}, 
\end{equation}
\begin{equation}
 e^{-\beta J \sigma_i \sigma_j } \rightarrow 
 \left( \begin{array}{cc} 
  e^{-\beta J}  &  e^{\beta J}  \\
 e^{\beta J}  &  e^{-\beta J} 
\end{array} \right)
= \sum_{m=1,2}V_{\sigma_i m}W_{\sigma_j m}, 
\end{equation}
where we find
\begin{equation}
 W = 
 \left( \begin{array}{cc} 
 \sqrt{\cosh(\beta J)} & \sqrt{\sinh(\beta J)}  \\
 \sqrt{\cosh(\beta J)} & -\sqrt{\sinh(\beta J)}
\end{array} \right)  ,
\label{eq:W}
\end{equation}
\begin{equation}
 V = 
 \left( \begin{array}{cc} 
 \sqrt{\cosh(\beta J)} & -\sqrt{\sinh(\beta J)}  \\
 \sqrt{\cosh(\beta J)} & \sqrt{\sinh(\beta J)}
\end{array} \right)  .
\label{eq:V}
\end{equation}
Using gauge fixing of $XX^{-1}=I$, we find that the change of $V \rightarrow W$ and $W \rightarrow V$
does not alter the partition function. Explicitly, we note that, for the antiferromagnetic case,
\begin{eqnarray}
\sum_{m=1,2}V_{\sigma_i m}W_{\sigma_j m} &=& \sum_{m=1,2}V_{\sigma_i m} \sum_{l,n=1,2}X_{m n}X^{-1}_{n l}W_{\sigma_j l} \nonumber \\
&=&  \sum_{n=1,2}\sum_{m=1,2}V_{\sigma_i m} X_{m n}\sum_{l=1,2}X^{-1}_{n l}W_{\sigma_j l} \nonumber \\
&=& \sum_{n=1,2}W_{\sigma_i n}V_{\sigma_j n}, 
\end{eqnarray}
where we use 
\begin{equation}
 X = 
 \left( \begin{array}{cc} 
 1 & 0  \\
 0 & -1
\end{array} \right)  .
\label{eq:W}
\end{equation}

\begin{figure}
\includegraphics[width= 8.0 cm]{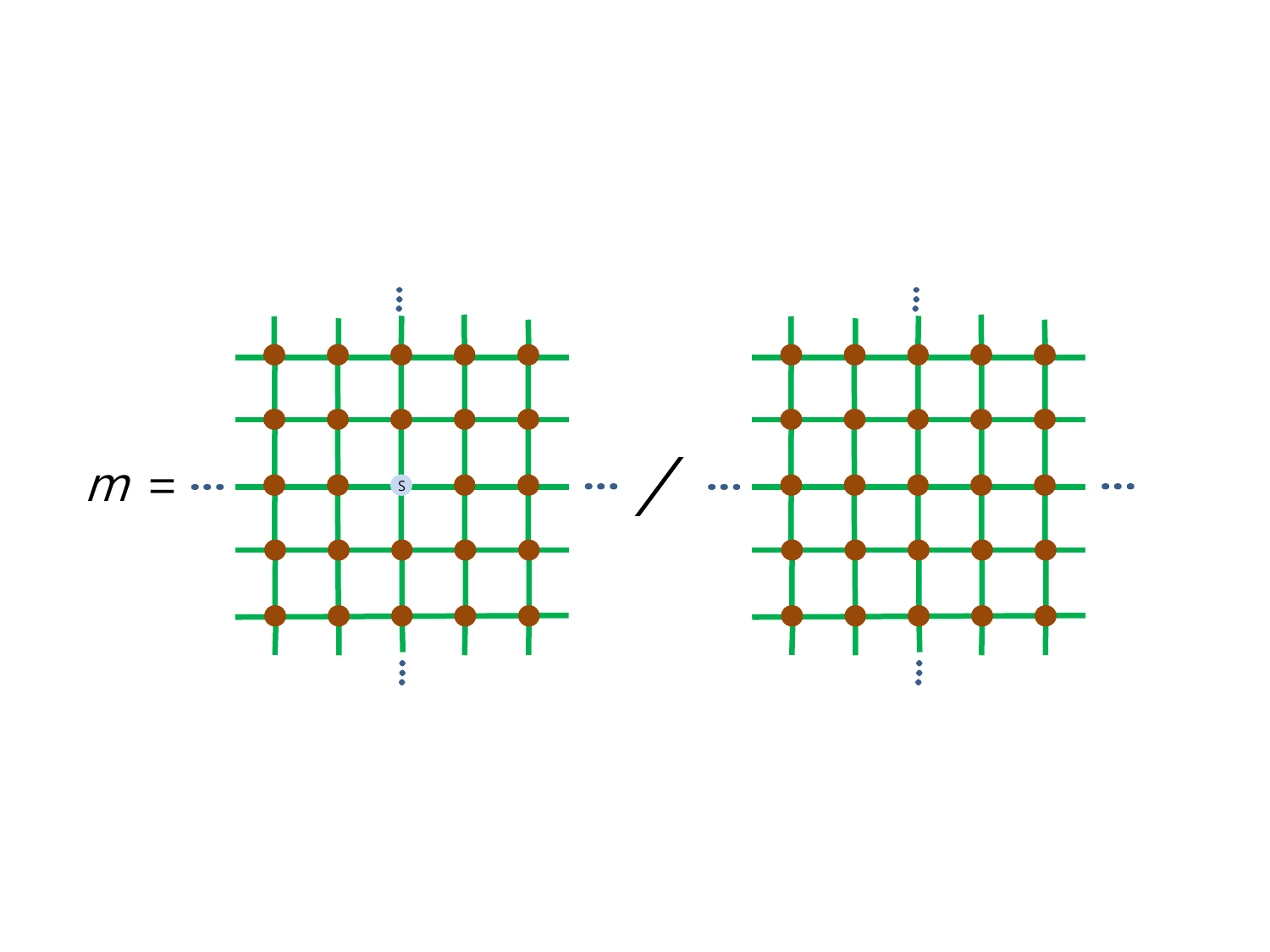}
\caption {Tensor network diagram for magnetization with normalization using partition function in the bulk limit with $\infty \times \infty$ lattice sites.
\label{fig:fig2}  }
\end{figure}

Note that the spins $\sigma_i$ and $\sigma_j$ are decoupled and connected 
via index $m$. This separation allows us to transform the partition function 
into a translation-invariant tensor network:
\begin{equation}
Z = \mbox{Tr} \prod_{i} T_{r_i u_i l_i d_i } ,
\label{eq:Z}
\end{equation}
where $i$ spans all lattice sites and $\mbox{Tr}$ must be the sum of all bond indices,
and the local tensor $T^{\mbox{f}}$ for the ferromagnetic model and $T^{\mbox{a}}$ for the antiferromagnetic model is defined as
\begin{equation}
T_{ruld}^{\mbox{f}}  = \sum_{\sigma = -1, 1}  e^{\beta B \sigma} W_{\sigma r}W_{\sigma u}W_{\sigma l}W_{\sigma d},
\label{eq:Tf}
\end{equation}
\begin{equation}
T_{ruld}^{\mbox{a}}  = \sum_{\sigma = -1, 1}  e^{\beta B \sigma} W_{\sigma r}V_{\sigma u}V_{\sigma l}W_{\sigma d},
\label{eq:Ta}
\end{equation}
which has $2 \times 2 \times 2 \times 2 = 16$ elements.

The four-legged constitutive tensor $T^{\mbox{f}}$ and $T^{\mbox{a}}$ is used to form the partition function as shown in Fig. \ref{fig:fig1}. It is emphasized that the tensor in the partition function is not unique due to gauge fixing. It is remarkable that the tensor $T^{\mbox{f}}_{ruld}$ exhibits index permutation symmetry while $T^{\mbox{a}}_{ruld}$ does not.

For magnetization, we consider the defect-like tensor written as
\begin{equation}
S_{ruld}^{\mbox{f}}  = \sum_{\sigma = -1, 1}  \sigma e^{\beta B \sigma} W_{\sigma r}W_{\sigma u}W_{\sigma l}W_{\sigma d} ,
\label{eq:Sf}
\end{equation}
\begin{equation}
S_{ruld}^{\mbox{a}}  = \sum_{\sigma = -1, 1}  \sigma e^{\beta B \sigma} W_{\sigma r}V_{\sigma u}V_{\sigma l}W_{\sigma d} ,
\label{eq:Sa}
\end{equation}
The value of the magnetization is given by the ratio of the network containing the defect-like tensor $S^{\mbox{f}}$ and $S^{\mbox{a}}$ divided by the partition function network as shown in Fig. \ref{fig:fig2}. 

\section{DMRG Algorithm}

In the framework of tensor network representation for the partition function, the main
task is to contract the network. 
The coarse-graining method to contract the network has a drawback
due to the defect-like tensor in an infinite system as shown in Fig. \ref{fig:fig2}.
Thus, we adopt iDMRG to contract the network here.

To contract the two dimensional tensor network of the partition function, we introduce the boundary state living on the boundary. Then, we 
consider the eigenvalue equation, which is written
in terms of the fundermental operator $T$ such as
\begin{eqnarray}
\begin{array}{c} \\
\cdots \\
\end{array}
\begin{array}{c} \\
- \\ 
  \\
- \\
\end{array}
\begin{array}{c} | \\
T \\
 | \\
A \\
\end{array}
\begin{array}{c} \\
- \\ 
  \\
- \\
\end{array}
\begin{array}{c} | \\
T \\
| \\
B \\
\end{array}
\begin{array}{c} \\
- \\ 
  \\
- \\
\end{array}
\begin{array}{c} | \\
T \\
| \\
A \\
\end{array}
\begin{array}{c} \\
- \\ 
  \\
- \\
\end{array}
\begin{array}{c} | \\
T \\
 | \\
B \\
\end{array}
\begin{array}{c} \\
- \\ 
  \\
- \\
\end{array}
\begin{array}{c} \\
\\
\cdots \\  
\label{eq:MPS}
\end{array}  
\\ \nonumber 
\\ \nonumber 
\\ \nonumber 
=\lambda_{\mbox{max}} \times
\begin{array}{c} \\
\cdots \\
\end{array}
\begin{array}{c} \\
- \\
\end{array}
\begin{array}{c} | \\
A \\
\end{array}
\begin{array}{c} \\
- \\
\end{array}
\begin{array}{c} | \\
B \\
\end{array}
\begin{array}{c} \\
- \\
\end{array}
\begin{array}{c} | \\
A \\
\end{array}
\begin{array}{c} \\
- \\
\end{array}
\begin{array}{c} | \\
B \\
\end{array}
\begin{array}{c} \\
- \\
\end{array}
\begin{array}{c} \\
\cdots \\
\end{array}
\end{eqnarray}
where $\lambda_{\mbox{max}}$ should be the dominant eigenvalue. This one-dimensional quantum model defined by this eigenvalue 
equation corresponds the finite temperature two-dimensional statistical model.

\begin{figure}
\includegraphics[width= 8.0 cm]{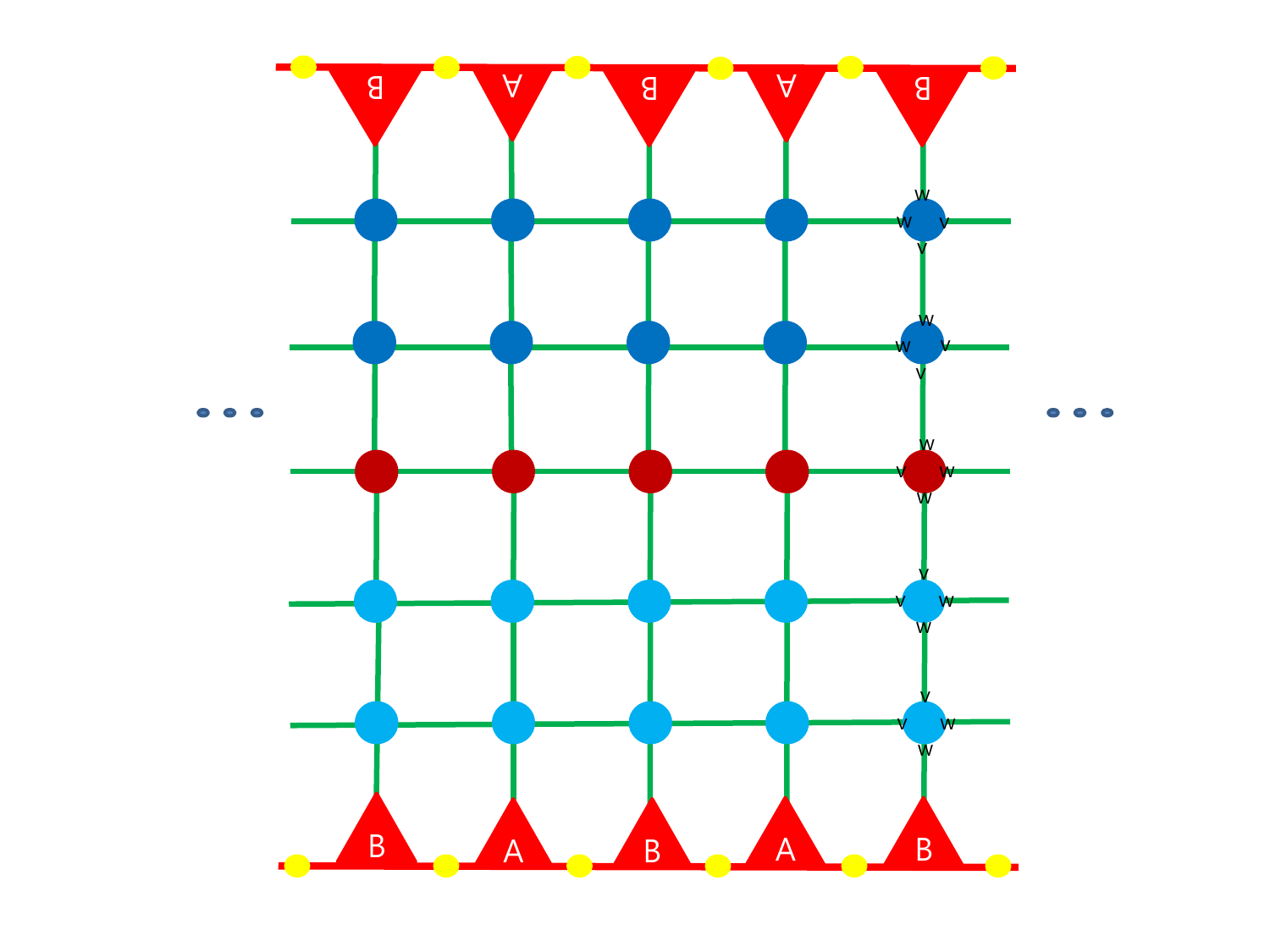}
\caption {Action of the matrix product operator on the matrix product state in the formulation of the row-to-row transfer matrix. 
The Schmidt coefficients between $A$ and $B$ are obtained by SVD decomposition.
Tensor contraction is achieved by this process. By simple gauge transformation, the upper contraction is the same as
the lower contraction.
\label{fig:fig3} }
\end{figure}

In the thermodynamic limit, when the row-to-row operator acts infinite times,
the boundary eigenstate with a dominant eigenvalue can only survive
as shown in Eq. (\ref{eq:MPS}).
Since the row-to-row transfer matrix plays the same role as MPO,
the boundary state is assumed to be MPS.
This special MPS will be determined by iDMRG with MPO where the element tensor operator
is given by the four-legged tensor $T$ as shown in Fig. \ref{fig:fig1}.
In the process of iDMRG,
we determine the three-legged tensors $A$, $B$ and the Schmidt coefficients $\lambda_{i}$
between $A$ and $B$.
For the ferromagnetic case, since $T^{\mbox{f}}_{ruld}$ has full index permutation symmetry,
the MPO multiplication process on MPS in the bottom-up, top-down, left-right 
or right-left directions gives us the same result.
Moreover, due to the symmetry, $A$ is identical to $B$ for the ferromagnetic case.
The fact that $A=B$ ensures that there are no odd-even effects in this system.
On the other hand, for the antiferromagnetic case,
since $T^{\mbox{a}}_{ruld}$ has a lower index permutation symmetry, one can
suspect a different entanglement entropy between the bottom-up process and the top-down process.
However, when the following gauge fixing is applied to upper half of the system, 
\begin{equation}
W_{\sigma r}V_{\sigma u}V_{\sigma l}W_{\sigma d}
\rightarrow
V_{\sigma r}W_{\sigma u}W_{\sigma l}V_{\sigma d},
\label{eq:gauge}
\end{equation}
it guarantees that the bottom-up process and the top-down process are exactly the same as shown 
in Fig. \ref{fig:fig3}.

After consistently optimizing the tensors and the coefficients, 
we are ready to compute the entanglement entropy and magnetization. 
To detect the critical points given by the partition function singularity, we use
the half-chain entanglement entropy defined by the Schmidt coefficients $\lambda_{i}$ such as
\begin{equation}
S_{h} = - \sum_{i=1}^{\chi} \lambda^{2}_{i}\log \lambda^{2}_{i}  ,
\label{eq:S}
\end{equation}
where $\chi$ is the environment bond dimension.

\section{Numerical Results: Magnetization and Entanglement Entropy}

The magnetization as a local observable can be evaluated by inserting the corresponding impurity tensor into the original tensor network for the partition function as shown in Fig. \ref{fig:fig2}. The actual evaluation of the denominator and numerator in Fig. \ref{fig:fig2} was done by iDMRG. After recursive iteration on iDMRG, all tensor elements are determined. The so-called left and right environment tensors in iDMRG are used in the magnetization evaluation.

\begin{figure}
\includegraphics[width= 8.0 cm]{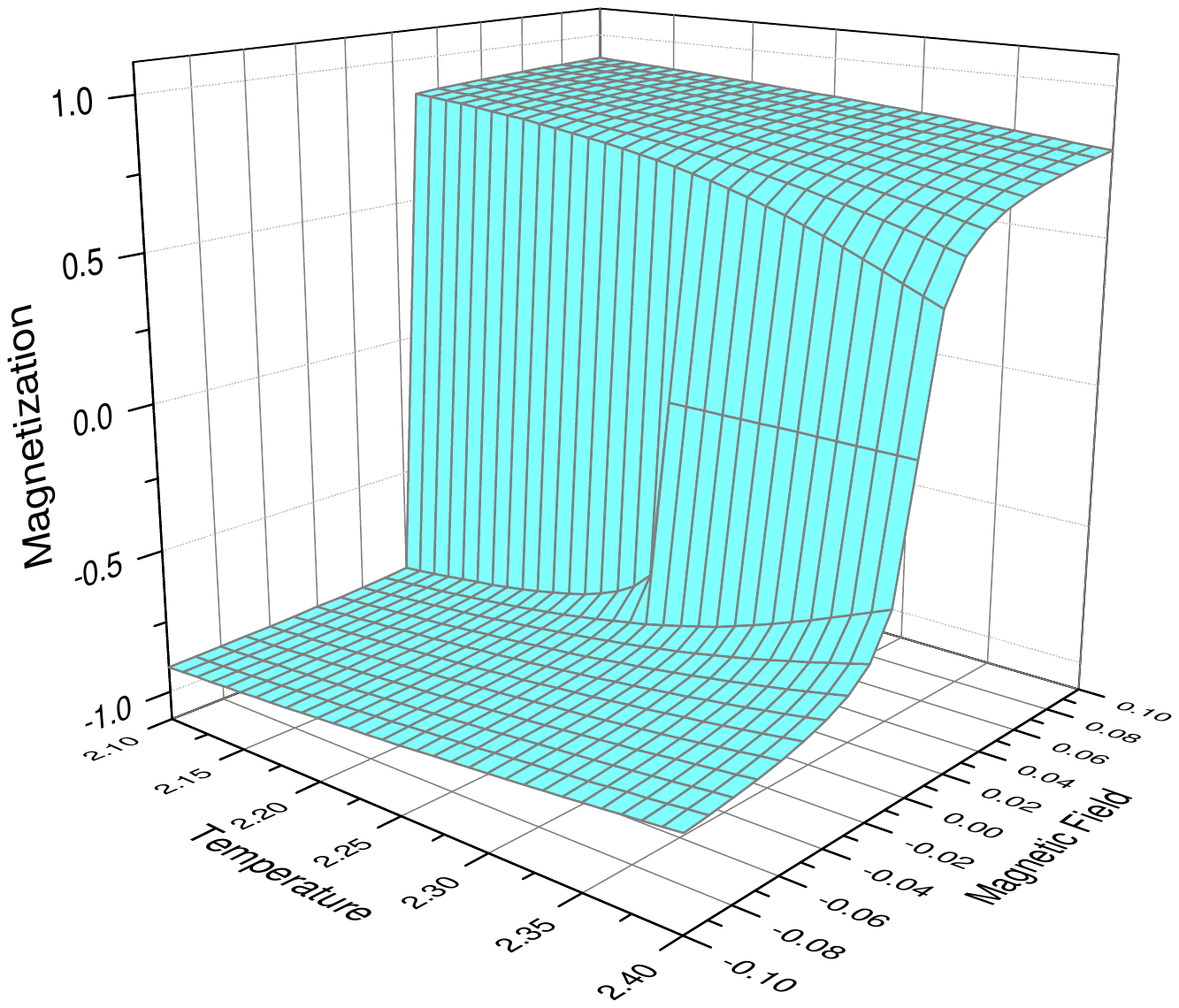}
\caption {Magnetization for the ferromagnetic Ising model.
The exact magnetization is found at zero magnetic field while
this numerical result can not be compared since the exact solution is not known
at non-zero magnetic field.
 \label{fig:fig4} }
\end{figure}

\begin{figure}
\includegraphics[width= 8.0 cm]{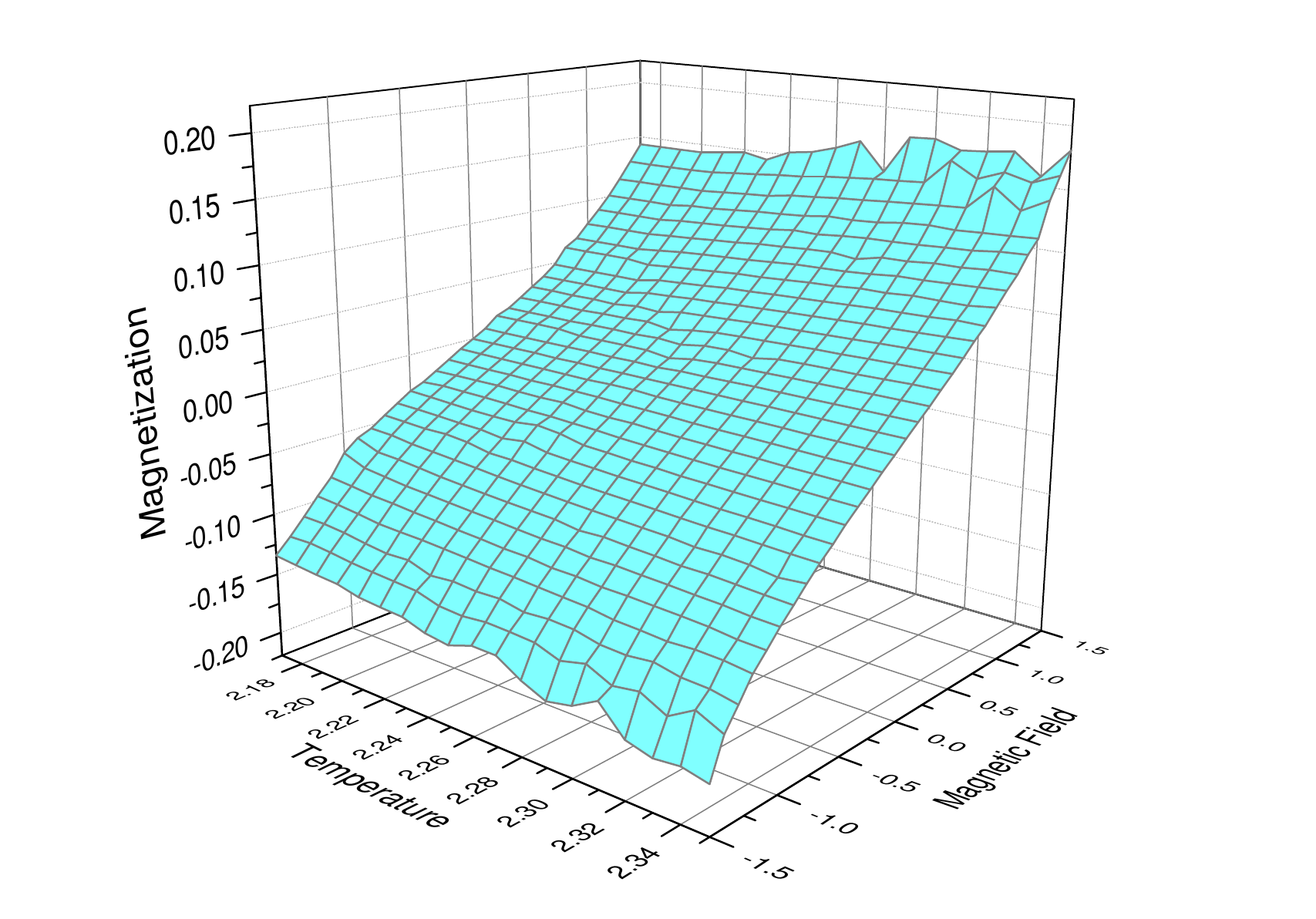}
\caption {Total magnetization for the antiferromagnetic Ising model.
The unstable data presented for a large magnetic field suggest that
we need a longer simulation time.
 \label{fig:fig5} }
\end{figure}

The critical temperature $T_{c}$ for the two-dimensional square lattice Ising model
with no external magnetic field
has been accurately calculated by Onsager \cite{Onsager}
as
\begin{equation}
T_{c} = \frac{2}{\ln (1 + \sqrt{2})} = 2.269185314\cdots ,
\label{eq:tc}
\end{equation}
This value gives us a good indication of the accuracy of our numerical results
obtained by simulation with input grid parameters of temperature and field
by fixing $J=1$ and $\chi = 60$.

\begin{figure}
\includegraphics[width= 8.0 cm]{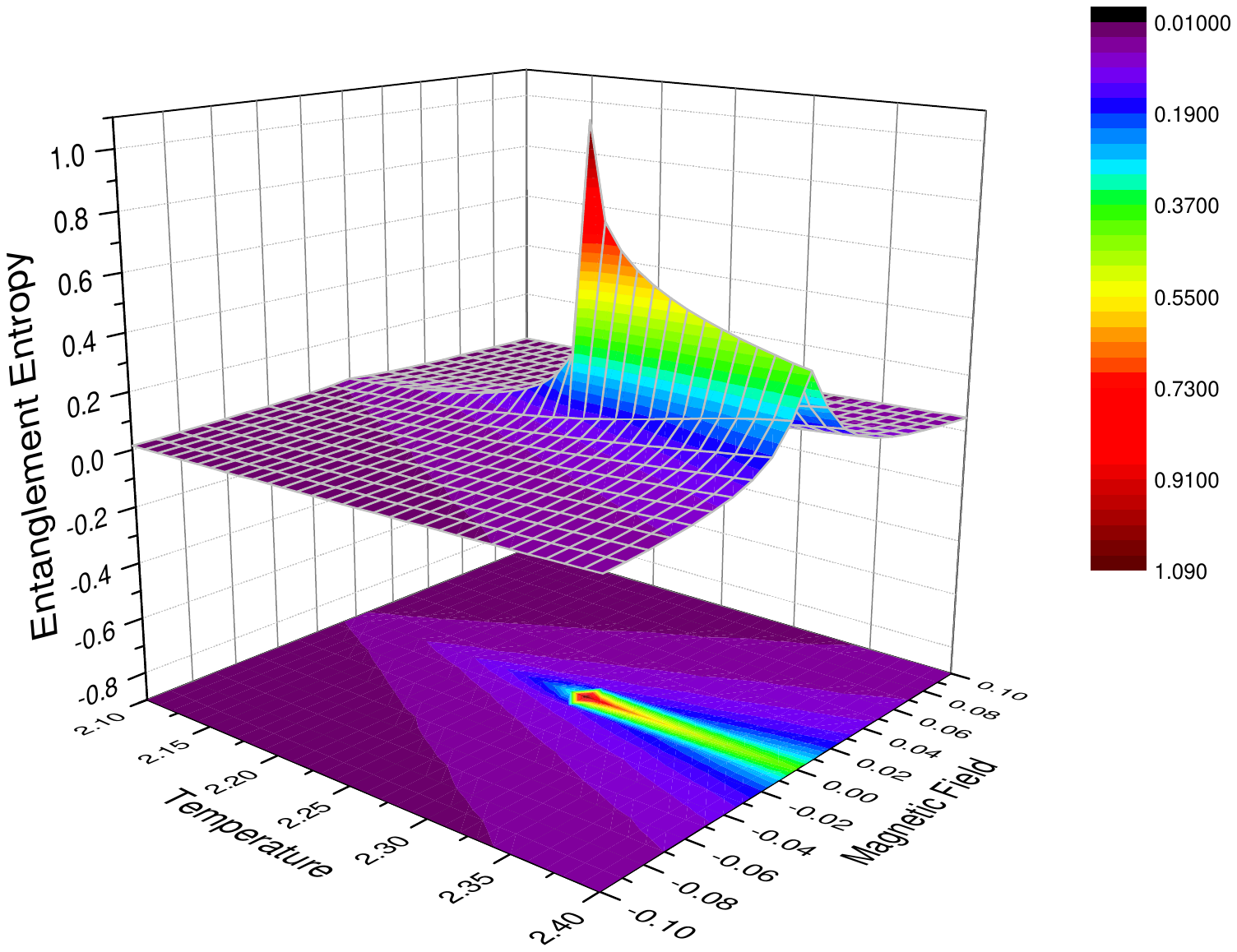}
\caption {
Half-chain entanglement entropy as a function of temperature and magnetic field
in the ferromagnetic Ising model. We notice a clear peak towards $T=2.27$ and zero field, which is the critical point. The entanglement entropy develops a singularity at $T_{c}$ 
by increasing the MPS bond dimension.
\label{fig:fig6} }
\end{figure}

We present the numerical result of the magnetization for the ferromagnetic case in Fig. \ref{fig:fig4} and for the antiferromagnetic case in Fig. \ref{fig:fig5}.
We find that there is a good agreement between our result at zero magnetic field and
the exact magnetization obtained by Yang \cite{CNYang} in the ferromagnetic case of Fig. \ref{fig:fig4}. For the antiferromagnetic case of Fig. \ref{fig:fig5} with zero magnetic field, we find that the magnetization vanishes beyond $T=2.27$ while the opposite magnetization is observed in a checkerboard pattern below $T=2.27$ keeping zero total magnetization.
In the antiferromagnetic case with nonzero magnetic fields, the total magnetization depends on the field strength. We note that the slope of the magnetization with respect to the magnetic field changes across the critical line forming the parabolic shape in the parameter space of temperature and magnetic field.

\begin{figure}
\includegraphics[width= 8.0 cm]{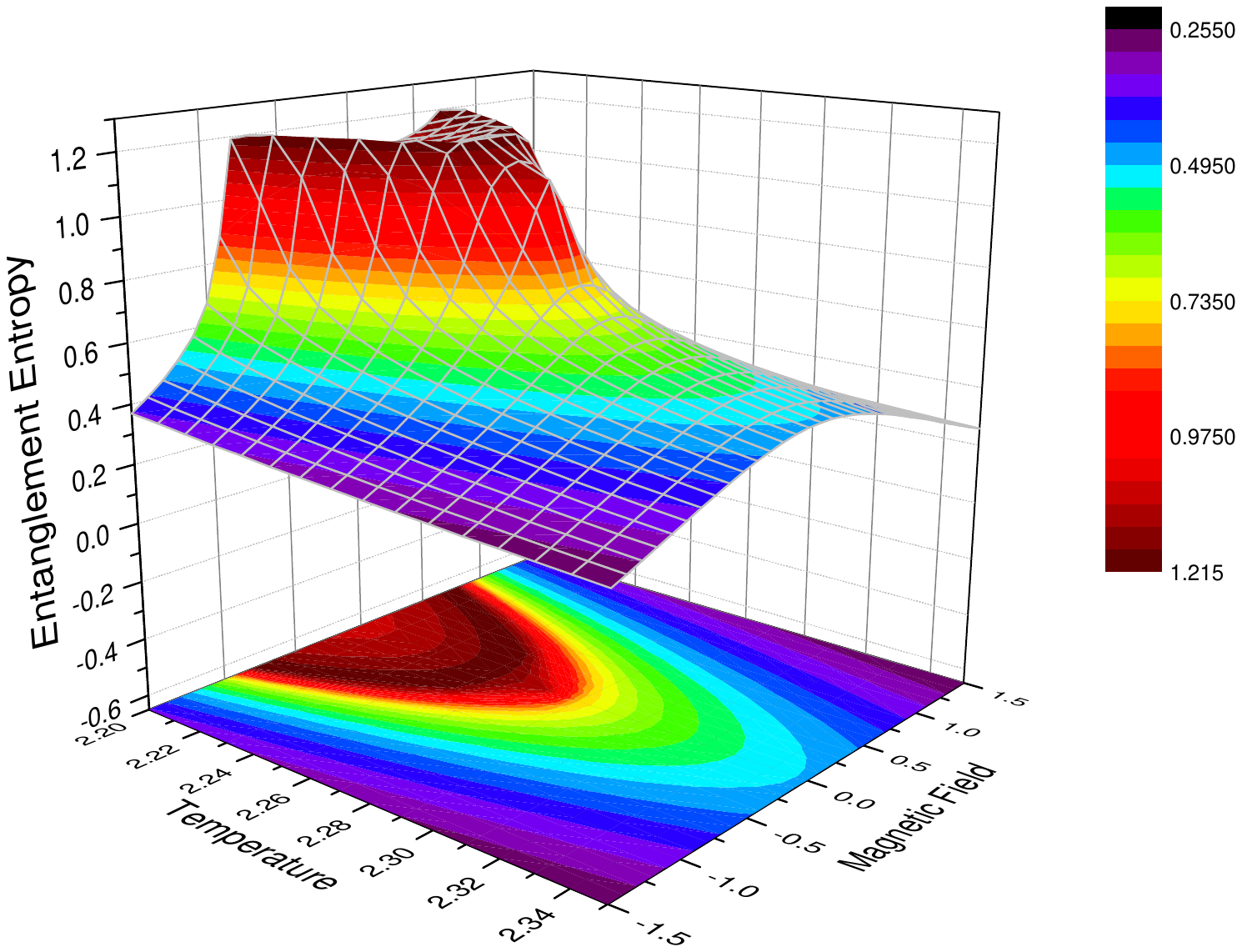}
\caption {
Half-chain entanglement entropy as a function of temperature and magnetic field
in the antiferromagnetic Ising model. We notice a parabolic region inside which we observe the high entropy. It is emphasized that the critical region is not a point but a two-dimensional parabola in the parameter space. \label{fig:fig7} }
\end{figure}

In the iDMRG process, we find the Schmidt coefficents $\lambda_{i}$ between the three-legged tensors $A$ and $B$ as shown in Eq. (\ref{eq:MPS}).
Using the Schmidt coefficients $\lambda_{i}$, we calculate the entanglement entropy $S_{h}$ in Eq. (\ref{eq:S}).
This entanglement entropy is the main numerical result. We present the half-chain entanglement entropy as a function of temperature and magnetic field
for ferromagnetic and antiferromagnetic cases as shown in Fig. \ref{fig:fig6} and Fig. \ref{fig:fig7}. We find the sharp peak at the point near $T=2.27$ and zero field for the ferromagnetic case, while there is a parabolic region where all values of 
the large entropy are almost the same for the antiferromagnetic case.
The parabolic boundary line defined by the entropy maxima is consistent with the result obtained by  M\"{u}ller-Hartmann and Zittartz \cite{Muller}.

\section{CONCLUSION}

In summary, we have used a tensor network method with iDMRG to find the critical region in the antiferromagnetic Ising model with a magnetic field. We highlight the role of gauge fixing in the construction of the partition function with tensor networks. The constitutive tensor of each lattice site exhibits complete symmetry for the ferromagnetic case while the tensor exhibits an asymmetric property for the antiferromagnetic case.
The partition function can be interpreted as the product of a one-dimensional transfer matrix operator.

To determine the partition function, we solve the eigenvalue equation defined by the one-dimensional transfer matrix operator. This eigenvalue problem provides us with a one-dimensional quantum analogue. The singularity of the entanglement entropy given by the quantum analogue shows a criterion for determining phase transitions 
in the classical two-dimensional statistical systems.
The corresponding entanglement entropy exhibits significantly distinct forms for the ferromagnetic and antiferromagnetic Ising model with a magnetic field.

We conclude that entanglement entropy plays the role of a marker indicating phase transitions.
Our method using entanglement entropy is useful for determining phase transitions,
provided that the partition function is written as a tensor network.

\begin{acknowledgments}
This work was partially supported by the Basic Science Research
Program through the National Research Foundation of Korea (NRF)
funded by the Ministry of Education, Science and Technology (Grant
No. NRF-2021R1F1A1052347). The author would like to thank P. K. Mohanty and D. Poilblanc
for helpful discussions.
\end{acknowledgments}

\end{document}